\documentclass[12pt]{article}
\usepackage{epsfig,graphicx}
\usepackage{fpcp04}

\def\to           {\ensuremath{\rightarrow}\xspace}
\def\Bztorhoprhom {\ensuremath{\Bz (\Bzb) \to \rho^+ \rho^- }\xspace}
\def\Bztorhozrhoz {\ensuremath{\Bz (\Bzb) \to \rho^0 \rho^0 }\xspace}
\def\Bztorhoprhoz {\ensuremath{\B^\pm \to \rho^\pm \rho^0 }\xspace}
\def\Bztorhopi    {\ensuremath{ \Bz (\Bzb) \to \rho^\pm \pi^\mp}\xspace}
\def\Bztopippim   {\ensuremath{\Bz (\Bzb) \to \pi^+ \pi^- }\xspace}

\def\Bztopippim {\ensuremath{\Bz (\Bzb) \to \pi^+ \pi^- }\xspace}
\def\Bztopizpiz {\ensuremath{\Bz (\Bzb) \to \pi^0 \pi^0 }\xspace}

\def\C       {\ensuremath{ C }}
\def\S       {\ensuremath{ S }}
\def\acp     {\ensuremath{ {\cal A}_{CP} }}
\def\clong   {\ensuremath{ C_{L} }}
\def\slong   {\ensuremath{ S_{L} }}
\def\ctran   {\ensuremath{ C_{T} }}
\def\stran   {\ensuremath{ S_{T} }}
\def\fL      {\ensuremath{ f_L }}
\def\ptrue   { \fL }
\def\deltat  { \ensuremath{\Delta t }}

\def\dalphaeff {\ensuremath{ \alpha_{eff} - \alpha } }

\def\br        {\ensuremath{ {\cal {B}}} }
\def\penguin   {\ensuremath{ { (\rm penguin) } } }

\def\beq{\begin{equation}}
\def\eeq#1{\label{#1}\end{equation}}
\def\eeqn{\end{equation}}

\def\beqa{\begin{eqnarray}}
\def\eeqa#1{\label{#1}\end{eqnarray}}
\def\eeqan{\end{eqnarray}}

\let\bar=\overbar

\def\Dslash{\not{\hbox{\kern-4pt $D$}}}
\def\dslash{\not{\hbox{\kern-2pt $\del$}}}

\def\msb{{\bar{\ssstyle M \kern -1pt S}}}

\def\BB0bar{B^0 {\overline B}^0}
\def\BB0dbar{B_d^0 {\overline B}_d^0}
\def\BB0sbar{B_s^0 {\overline B}_s^0}

\RequirePackage{xspace}

\usepackage{relsize}
\def\babar{\mbox{\slshape B\kern-0.1em{\smaller A}\kern-0.1em
    B\kern-0.1em{\smaller A\kern-0.2em R}}}

\def\epem       {\ensuremath{e^+e^-}\xspace}

\def\qqbar {\ensuremath{q\overline q}\xspace}

\def\piz   {\ensuremath{\pi^0}\xspace}

\def\pim   {\ensuremath{\pi^-}\xspace}

\def\Kbar  {\kern 0.2em\overline{\kern -0.2em K}{}\xspace}

\def\Kz    {\ensuremath{K^0}\xspace}
\def\Kzb   {\ensuremath{\Kbar^0}\xspace}
\def\KzKzb {\ensuremath{\Kz \kern -0.16em \Kzb}\xspace}
\def\Kp    {\ensuremath{K^+}\xspace}
\def\Km    {\ensuremath{K^-}\xspace}

\def\KpKm  {\ensuremath{\Kp \kern -0.16em \Km}\xspace}

\def\Dbar    {\kern 0.2em\overline{\kern -0.2em D}{}\xspace}

\def\Dz      {\ensuremath{D^0}\xspace}
\def\Dzb     {\ensuremath{\Dbar^0}\xspace}
\def\DzDzb   {\ensuremath{\Dz {\kern -0.16em \Dzb}}\xspace}
\def\Dp      {\ensuremath{D^+}\xspace}
\def\Dm      {\ensuremath{D^-}\xspace}

\def\DpDm    {\ensuremath{\Dp {\kern -0.16em \Dm}}\xspace}

\def\B       {\ensuremath{B}\xspace}
\def\Bbar    {\kern 0.18em\overline{\kern -0.18em B}{}\xspace}

\def\BB      {\ensuremath{B\Bbar}\xspace} 
\def\Bz      {\ensuremath{B^0}\xspace}
\def\Bzb     {\ensuremath{\Bbar^0}\xspace}
\def\BzBzb   {\ensuremath{\Bz {\kern -0.16em \Bzb}}\xspace}
\def\Bu      {\ensuremath{B^+}\xspace}
\def\Bub     {\ensuremath{B^-}\xspace}
\def\Bp      {\ensuremath{\Bu}\xspace}

\def\BpBm    {\ensuremath{\Bu {\kern -0.16em \Bub}}\xspace}

\mathchardef\Upsilon="7107
\def\Y#1S{\ensuremath{\Upsilon{(#1S)}}\xspace}

\mathchardef\Deltares="7101
\mathchardef\Xi="7104
\mathchardef\Lambda="7103
\mathchardef\Sigma="7106
\mathchardef\Omega="710A

\def\Deltabar{\kern 0.25em\overline{\kern -0.25em \Deltares}{}\xspace}
\def\Lbar{\kern 0.2em\overline{\kern -0.2em\Lambda\kern 0.05em}\kern-0.05em{}\xspace}
\def\Sigbar{\kern 0.2em\overline{\kern -0.2em \Sigma}{}\xspace}
\def\Xibar{\kern 0.2em\overline{\kern -0.2em \Xi}{}\xspace}
\def\Obar{\kern 0.2em\overline{\kern -0.2em \Omega}{}\xspace}
\def\Nbar{\kern 0.2em\overline{\kern -0.2em N}{}\xspace}
\def\Xb{\kern 0.2em\overline{\kern -0.2em X}{}\xspace}

\def\Bztorhopi  {\ensuremath{\Bz \to \rho^+ \pim}\xspace}
\def\Bztorhorho {\ensuremath{\Bz \to \rho \rho}\xspace}

\def\mes        {\mbox{$m_{\rm ES}$}\xspace}

\def\DeltaE     {\mbox{$\Delta E$}\xspace}

\newcommand{\tev}{\ensuremath{\mathrm{\,Te\kern -0.1em V}}\xspace}
\newcommand{\gev}{\ensuremath{\mathrm{\,Ge\kern -0.1em V}}\xspace}
\newcommand{\mev}{\ensuremath{\mathrm{\,Me\kern -0.1em V}}\xspace}
\newcommand{\kev}{\ensuremath{\mathrm{\,ke\kern -0.1em V}}\xspace}
\newcommand{\ev}{\ensuremath{\mathrm{\,e\kern -0.1em V}}\xspace}
\newcommand{\gevc}{\ensuremath{{\mathrm{\,Ge\kern -0.1em V\!/}c}}\xspace}
\newcommand{\mevc}{\ensuremath{{\mathrm{\,Me\kern -0.1em V\!/}c}}\xspace}
\newcommand{\gevcc}{\ensuremath{{\mathrm{\,Ge\kern -0.1em V\!/}c^2}}\xspace}
\newcommand{\mevcc}{\ensuremath{{\mathrm{\,Me\kern -0.1em V\!/}c^2}}\xspace}

\def\mus  {\ensuremath{\rm \,\mus}\xspace}

\def\mus        {\ensuremath{\,\mu{\rm s}}\xspace}    

\def\to                 {\ensuremath{\rightarrow}\xspace}

\newcommand{\stat}{\ensuremath{\mathrm{(stat)}}\xspace}
\newcommand{\syst}{\ensuremath{\mathrm{(syst)}}\xspace}

\def\pep2{PEP-II}

\def\gsim{{~\raise.15em\hbox{$>$}\kern-.85em
          \lower.35em\hbox{$\sim$}~}\xspace}
\def\lsim{{~\raise.15em\hbox{$<$}\kern-.85em
          \lower.35em\hbox{$\sim$}~}\xspace}

\def\CP                {\ensuremath{C\!P}\xspace}

\def\deltaz{\ensuremath{{\rm \Delta}z}\xspace}
\def\deltat{\ensuremath{{\rm \Delta}t}\xspace}
\def\deltamd{\ensuremath{{\rm \Delta}m_d}\xspace}

\xspace

\newcommand{\jprlBase}       {Phys.\ Rev.\ Lett.\xspace}
\newcommand{\jprBase}        {Phys.\ Rev.\xspace}
\newcommand{\jplBase}        {Phys.\ Lett.\xspace}

\newcommand{\plb}       [1]  {\jplBase\ B~{\bf #1}}

\newcommand{\jprl}      [1]  {\jprlBase\ {\bf #1}}
\newcommand{\jprd}      [1]  {\jprBase\ D~{\bf #1}}

\newcommand{\progtp}    [1]  {{Prog.\ Th.\ Phys.\ {\bf #1}}}

\def\jetset74   {\mbox{\tt Jetset \hspace{-0.5em}7.\hspace{-0.2em}4}\xspace}

\begin{document}

\begin{flushright}
\babar-PROC-04/138\\
SLAC-PUB-10874\\
\end{flushright}

\Title{Measurements of $\sin2\alpha/\phi_2$ from $B\to\pi\pi$,
$\rho \pi$ and $\rho\rho$ modes.}
\bigskip

%
\label{AJBevanStart}

%
\author{ Adrian Bevan\index{Bevan, A. J.} }

%
\address{Department of Physics\\
Liverpool University\\
Liverpool, United Kingdom\\
(from the \babar\ Collaboration.)\\
Talk given at Flavor Physics and \CP Violation, $4^{th}-9^{th}$ October 2004, Daegu, Korea.\\
}

\makeauthor\abstracts { B meson decays involving $b \to u$
transitions are sensitive to the unitarity triangle angle $\alpha$
(or $\phi_2$).  The \babar\ and Belle experiments have studied
B-meson decays to $\pi\pi$, $\rho\pi$ and $\rho\rho$ final states.
It is possible to combine these measurements to constrain $\alpha$
with a precision of ${\cal O} (10^\circ)$ and a central value of
approximately 100 degrees.  These results are consistent with
Standard Model expectations. }

\section{Introduction}

This is a summary of the experimental constraint on $\alpha$
obtained by the B-factories from B meson decays involving $b \to
u$ transitions. The final states of interest for these studies are
$B \to hh$, where $h = \rho,\pi$.  \CP Violation (CPV) in the
Standard Model of particle physics (SM) is described by a single
complex phase.  Interference between direct decay and decay after
\Bz-\Bzb mixing results in a time-dependent decay-rate asymmetry
between \Bz and \Bzb decaying to $h^+h^-$ that is sensitive to the
CKM~\cite{bevan:CKM} angle $\alpha \equiv
\arg\left[-V_{td}^{}V_{tb}^{*}/V_{ud}^{}V_{ub}^{*}\right]$. The
presence of loop (penguin) contributions introduces additional
phases that can shift the experimentally measurable parameter
$\alpha_{\mathrm{eff}}$ away from the value of $\alpha$.  In the
presence of penguin contributions $\alpha_{\mathrm{eff}} = \alpha
+ \delta\alpha_{\mathrm{penguin}}$. The CKM angle $\beta$ is well
known and consistent with SM predictions. Any constraint on
$\alpha$ constitutes a new test of the SM description of quark
mixing and CPV in B-meson decays.  Any significant deviation from
SM expectation would be a clear indication of new physics. The SM
prediction for $\alpha$ are $[95 \pm 9]^\circ$ from
\cite{bevan:utfitter} and $[98 \pm 16]^\circ$~\cite{bevan:ckmfitter}.

There are two classes of measurement that the B-factories are
pursuing, the main goal is to use SU(2) relations to relate
different $hh$ final states and limit
$\delta\alpha_{\mathrm{penguin}}$ in each of the modes
$\Bztopippim$, $\Bztorhopi$ and $\Bztorhoprhom$.  In some cases
only weak constraints on $\alpha$ are obtained when using SU(2)
analysis and one must use a model dependent approach to obtain a
significant result.

Experimental results are quoted with statistical errors preceding
systematic errors unless otherwise stated.

\section{Isospin analysis of $B\to hh$}

One can relate the amplitudes of $B$ decays to $\pi\pi$ final
states by noting that $\frac{1}{\sqrt{2}}A^{+-}=A^{+0}-A^{00}$ and
$\frac{1}{\sqrt{2}}\overline{A}^{+-}=\overline{A}^{+0}-\overline{A}^{00}$~\cite{bevan:gronaulondon}.
These two relations correspond to triangles in a complex plane. In
the absence of Electroweak penguin contributions,
$|A^{+0}|=|\overline{A}^{+0}|$, i.e. the two triangles have a
common base. The phase difference between $A^{+-}$ and
$\overline{A}^{+-}$ is related to $|\dalphaeff|$.  Thus, in order
to measure $\alpha$ one must measure the branching fractions (\br)
and charge asymmetries (\acp) of $B$ decays to $\pi^+\pi^-$,
$\pi^\pm\pi^0$, $\pi^0\pi^0$. The decays of $B\to\rho\rho$ and
$B\to\rho\pi$ are more complicated.  In the case of the former
decays one has up to three isospin triangle analyses to perform to
extract maximal information from data.  The final state of $B$
decays to $\rho^\pm\pi^\mp$ is not a \CP eigenstate, which results
in the need either for a pentagon isospin analysis~\cite{bevan:lipkin}
or a Dalitz Plot (DP) analysis~\cite{bevan:sniderquinn}.

For initial measurements of $\alpha$, one can safely neglect
electroweak penguin contributions to the isospin analysis as these
are a small correction to
$\alpha$~\cite{bevan:electroweak_pengin_calculation}. Some estimates of
the effect of SU(2) symmetry breaking
exist~\cite{bevan:isospin_symmetry_breaking}, where any correction is
estimated to be much less than $10^\circ$.

\section{Experimental techniques}

Continuum $\epem \to \qqbar$ ($q = u,d,s,c$) events are the
dominant background. To discriminate signal from continuum one
uses either a neural network, Fisher discriminant or likelihood
ratio of event shape variables.  Signal candidates ($B_{rec}$) are
identified kinematically using two variables, the difference
\DeltaE between the CM energy of the \B candidate and
$\sqrt{s}/2$, and the beam-energy substituted mass $\mes =
\sqrt{(s/2 + {\mathbf {p}}_i\cdot {\mathbf {p}}_B)^2/E_i^2-
{\mathbf {p}}_B^2}$, where $\sqrt{s}$ is the total CM energy. The
\B momentum ${\mathbf {p}_B}$ and four-momentum of the initial
state $(E_i, {\mathbf {p}_i})$ are defined in the laboratory
frame.  The signal is extracted using an extended unbinned maximum
likelihood fit.

To study the time-dependent asymmetry one needs to measure the
proper time difference, \deltat, between the two \B\ decays in the
event, and to determine the flavor tag of the other \B-meson
($B_{\rm tag}$). The time difference between the decays of the two
neutral $B$ mesons in the event ($B_{\rm rec}$, $B_{\rm tag}$) is
calculated from the measured separation \deltaz between the
$B_{\rm rec}$ and $B_{\rm tag}$ decay vertices. The $B_{\rm rec}$
vertex is determined from the two charged-pion tracks in its
decay. The $B_{\rm tag}$ decay vertex is obtained by fitting the
other tracks in the event.

The signal decay-rate distribution $f_+ (f_-)$ for $B_{\rm tag}$=
\Bz (\Bzb) is given by:
\begin{eqnarray*}
f_{\pm}(\deltat) = \frac{e^{-\left|\deltat\right|/\tau}}{4\tau} [1
\pm S\sin(\deltamd\deltat) \mp \C\cos(\deltamd\deltat)]\,,
\nonumber
\end{eqnarray*}
where $\tau$ is the mean \Bz lifetime, \deltamd is the \Bz-\Bzb
mixing frequency.  For $B$ decays to $\rho^+\rho^-$, $S$= \slong\
or \stran\ and $C$= \clong\ or \ctran\ are the \CP asymmetry
parameters for the longitudinal and transversely polarized signal.
The decay-rate distribution for $B$ decays to the
$\rho^\pm\pi^\mp$ final state is described below.  The fitting
function takes into account mistag dilution and is convoluted with
the \deltat resolution function.

CPV is probed by studying the time-dependent decay-rate asymmetry
${\cal A} = (
  R (\deltat) - \overline{R}(\deltat) ) / (  R (\deltat) + \overline{R}(\deltat) )$,
  where $R$($\overline{R}$) is the decay-rate for \Bz(\Bzb) tagged
events.  This asymmetry has the form ${\cal
A}=S\sin(\deltamd\deltat)-\C\cos(\deltamd\deltat)$\footnote{Belle
use a different convention with $\C=-\acp$}. In the case of
charged B meson decays (and $\pi^0\pi^0$ as there is no vertex
information) one can study a time integrated charge asymmetry,
$\acp=(\overline{N}-N)/(\overline{N}+N)$, where $N$
($\overline{N}$) is the number of B ($\overline {\B}$) decays to
the final state.  A non-zero measurement of \S, \C\ or \acp\ for
any of the decays understudy would be a clear indication of \CP
violation.

\section{$B \to \pi\pi$}

The simplest decays to study in the pursuit of $\alpha$ are $B \to
\pi \pi$.  Both experiments have measurements for $B \to
\pi^+\pi^-$, $\pi^+\pi^0$ and $\pi^0\pi^0$
decays~\cite{bevan:babar_pipi,bevan:belle_pipi}.  These results are summarised
in Table~\ref{tbl:AJBevan_pipidata}.  All of these modes are now
well established experimentally and so it is possible to perform
an isospin analysis.

\begin{table}[!h]
\begin{center}
\begin{tabular}{c|cccc}
mode           & Expt   & \br & \S                        & \C \\
\hline
$\pi^+\pi^-$   & \babar & $4.7\pm 0.6\pm 0.2$ & $-0.30 \pm 0.17 \pm 0.03$ & $-0.09 \pm 0.15 \pm 0.04$\\
               & Belle  & $4.4\pm 0.6\pm 0.3$ & $-1.00 \pm 0.21 \pm 0.07$ & $-0.58 \pm 0.15 \pm
               0.07$\\ \hline
\end{tabular}
\begin{tabular}{c|cccc}
mode           & Expt   & \br                         & \acp\\
\hline
$\pi^\pm\pi^0$ & \babar & $5.8\pm 0.6 \pm 0.4$        &  $-0.01 \pm 0.10 \pm 0.02$  \\
               & Belle  & $5.0 \pm 1.2 \pm 0.5$       &  $-0.02 \pm 0.10 \pm 0.01$ \\
$\pi^0\pi^0$   & \babar & $1.17\pm 0.32 \pm 0.10$     &  $-0.12 \pm 0.56 \pm 0.06$ \\
               & Belle  & $2.32^{+0.44}_{-0.48}\,^{+0.22}_{-0.18}$  &  $-0.43 \pm 0.51^{+0.17}_{-0.16}$\\

\end{tabular}
\end{center}
\caption{Measurements of the $B \to \pi\pi$ decays.  BRs are in
units of $10^{-6}$.} \label{tbl:AJBevan_pipidata}
\end{table}

\begin{figure}[!h]
\begin{center}
\resizebox{8cm}{!}{\includegraphics{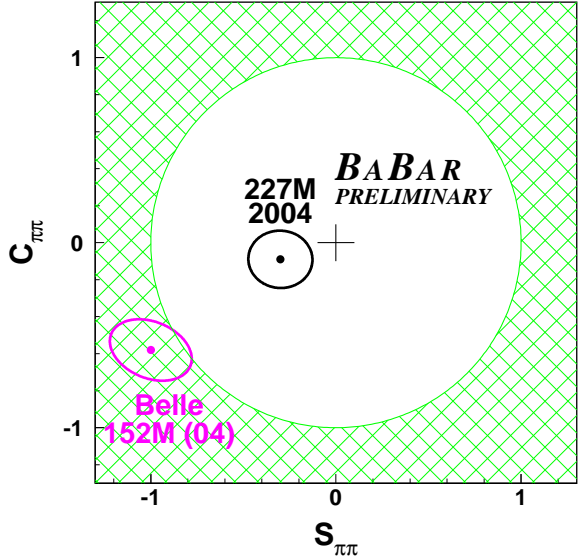}}
\end{center}
 \caption{A plot of the $S_{\pi\pi}$ vs $C_{\pi\pi}$ plane for the results from \babar\ and Belle.  The hatched region is unphysical.}
\label{fig:bevan_sc_pipi}
\end{figure}

One should note that the Belle measurement of \Bztopippim\
constitutes an observation of mixing-induced CPV in this decay at
a level of $5.2\sigma$, and evidence for direct CPV at a level of
$3.2\sigma$.  This is the second observation of CPV in B-meson
decays and is illustrated in Figure~\ref{fig:bevan_sc_pipi} . The
\babar\ data do not support this conclusion, and the discrepancy
between the two experiments is more than $3\sigma$. One must wait
for a significant increase in statistics from both experiments to
see if the results of the experiments become more compatible.

The $\pi\pi$ isospin analysis is limited by the value of
$\br(\Bztopizpiz)$.  Unfortunately this is neither large enough to
provide sufficient statistics with the current data set to perform
a precision measurement of \dalphaeff, nor small enough to enable
a strong bound on this quantity.  \babar\ have performed an
isospin analysis resulting in $|\dalphaeff|<35^\circ$ (90\% C.L.).
It is possible to derive model dependent constraints on $\alpha$,
as done by the Belle Collaboration;
$90<\alpha<146^\circ$~\cite{bevan:belle_pipi} (90\% C.L.) using
\cite{bevan:gronaurosner}.

\section{$B \to \rho\rho$}

The decays of $B \to \rho\rho$ correspond to a spin zero particle
decaying into two spin one, vector (V), particles.  As a result,
the \CP analysis of $B$ decays to $\rho^+\rho^-$ is complicated by
the presence of three helicity states ($h=0,\pm 1$). The $h=0$
state corresponds to longitudinal polarization and is \CP-even,
while neither the $h=+1$ nor the $h=-1$ state is an eigenstate of
\CP. The longitudinal polarization fraction $f_L$ is defined as
the fraction of the helicity zero state in the decay. The angular
distribution is
\begin{eqnarray}
&&\frac{d^2\Gamma}{\Gamma d\cos\theta_1 d\cos\theta_2}=
\frac{9}{4}\left(f_L \cos^2\theta_1 \cos^2\theta_2 + \frac{1}{4}(1-f_L) \sin^2\theta_1 \sin^2\theta_2 \right)
\end{eqnarray}
where $\theta_{i}, i=1,2$ is defined for each $\rho$ meson as the
angle between the \piz momentum in the $\rho$ rest frame and the
flight direction of the $B^0$ in this frame. The angle between the
$\rho$-decay planes is integrated over to simplify the analysis. A
full angular analysis of the decays is needed in order to separate
the definite \CP contributions; if however a single \CP channel
dominates the decay (which has been experimentally verified), this
is not necessary~\cite{bevan:Dunietz}. The longitudinal polarization
dominates this decay~\cite{bevan:Suzuki,bevan:my_rhorhoprl}. Not all of the
$\rho\rho$ final states have been observed, however as $\br(B^0
\to \rho^0\rho^0)$ is small, one can conservatively assume that is
longitudinally dominated when performing an isospin analysis.
There are assumptions in addition to SU(2) that are made when
performing a $\rho\rho$ isospin analysis. These are (i) neglect
Electroweak penguin contributions [a few $\,^\circ$ effect] (ii)
ignore possible $I=1$ components in the
decay~\cite{bevan:falk}\footnote{This is a reasonable assumption given
available statistics.}. The measured branching fractions, \ptrue,
\S\ and \C\ for $\B \to \rho\rho$ are summarised in Table
\ref{tbl:AJBevan_rhorhodata}.

\begin{table}[!h]
\begin{center}
\begin{tabular}{c|cccc}
mode & \br & \ptrue & \S & \C \\ \hline
$\rho^+\rho^-$   & $30 \pm 5 \pm 4$  & $0.99\pm 0.03 \pm 0.04$ & $-0.19 \pm 0.33 \pm 0.11$ & $-0.23 \pm 0.24 \pm 0.14$  \\
\end{tabular}

\begin{tabular}{c|cccc}
mode & Expt & \br & \ptrue & \acp \\ \hline
$\rho^\pm\rho^0$ & \babar & $22.5^{+5.7}_{-5.4}\pm 5.8$  & $0.^{+0.05}_{-0.07}$  & $-0.19 \pm 0.23 \pm 0.03$  \\
$\rho^\pm\rho^0$ & Belle  & $31.7 \pm 7.1^{+3.8}_{-6.7}$ & $0.^{+0.05}_{-0.07}$  & $0.00 \pm 0.22 \pm 0.03$ \\
$\rho^0\rho^0$   & \babar & $<1.1$ (90\% C.L.)& -                       & -  \\
\end{tabular}
\end{center}
\caption{Measurements of the $B \to \rho\rho$ decays.  BRs are in
units of $10^{-6}$.  \ptrue=1 is assumed in order to calculate the
upper limit on $\br(\Bztorhozrhoz)$.}
\label{tbl:AJBevan_rhorhodata}
\end{table}

Recent measurements of the $\Bp \to \rho^+\rho^0$ branching
fraction and upper limit for $\Bz \to \rho^0 \rho^0$
\cite{bevan:recentrhorho} indicate small penguin contributions in $\B
\to \rho \rho$, as has been found in some
calculations~\cite{bevan:aleksan}.  The decay mode $\Bztorhoprhoz$ has
only been studied using a small fraction of the available data of
the B-factories.  As has been remarked, it is important to see
this mode updated in the near future~\cite{bevan:utfitter,bevan:ckmfitter}.

\begin{figure}[!h]
\begin{center}
\resizebox{8cm}{!}{\includegraphics{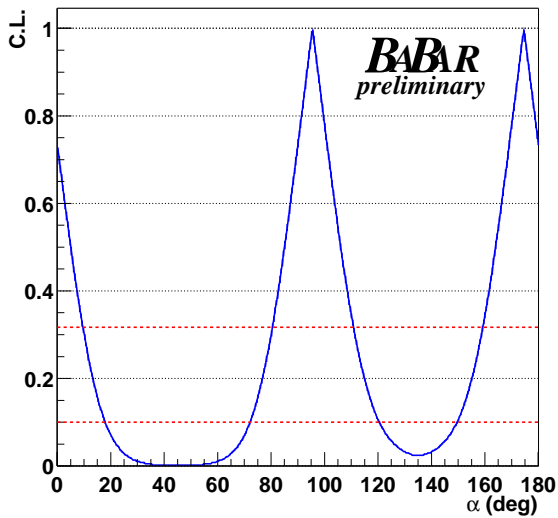}}
\end{center}
 \caption{A plot of the confidence level of $\alpha$ determined from \babar's $\rho\rho$ isospin analysis.}
\label{fig:bevan_rhorho_isospin}
\end{figure}

Given that penguin pollution is small, it is possible to perform
an isospin analysis of the longitudinal polarisation of the
$B\to\rho\rho$ decays, and use the results of \babar's
time-dependent \CP analysis of $\rho^+\rho^-$ \cite{bevan:my_rhorhoprl}
to constrain $\alpha$.  If one does this, using the
afforementioned assumptions, one obtains $\alpha=(96\pm 10 \stat
\pm 5 \syst \pm 11 \penguin)^\circ$.  The "penguin" error is
determined primarily from the experimental knowledge of the
$\rho^0\rho^0$ and $\rho^\pm\rho^0$ branching ratios.
Figure~\ref{fig:bevan_rhorho_isospin} shows the confidence level
plot of $\alpha$ corresponding to the isospin analysis of the
longitudinally polarized $\rho\rho$ data.

\section{$B \to \rho\pi$}

The decays $B \to \rho\pi$ can be analysed in two different ways.
The more straight forward approach is to cut away interference
regions of the $\pi^+\pi^-\pi^0$ DP and analyse the regions in the
vicinity of $\rho$ resonances.  This is called the Quasi-2-body
approach (Q2B) and it avoids the need to understand the
interference regions. The drawback of the Q2B method is that one
looses statistical power by cutting on the DP.  A corollory of
this is that one requires more statistics than the B-factories
currently have in order to get a significant constraint from the
pentagon isospin analysis. The alternative is to perform an
analysis of the $B\to \pi^+\pi^-\pi^0$ DP, accounting for the
interference between intersecting $\rho$ resonance bands and other
resonant structure. Both of these approaches have been studied by
\babar ~\cite{bevan:oldrhopi}. Belle has performed a Q2B analysis of
$\rho\pi$~\cite{bevan:bellerhopi}, and the first time-dependent DP
analysis has been performed by \babar~\cite{bevan:rhopidp}.

In the Q2B approach, one fits a time-dependence of
\begin{eqnarray*}
f_{\pm}(\deltat) = (1\pm
\acp)\frac{e^{-\left|\deltat\right|/\tau}}{4\tau} [ (S \pm \Delta
S \sin(\deltamd\deltat) - (C \pm \Delta C)
\cos(\deltamd\deltat)]\,, \nonumber
\end{eqnarray*}
where there are three additional parameters, a charged asymmetry,
\acp, between decays to $\rho^+\pi^-$ and $\rho^-\pi^+$ final
states, and two dilution parameters; $\Delta S$ and $\Delta C$.
The details of the DP analysis can be found in~\cite{bevan:rhopidp},
where one varies a larger number of parameters in the nominal fit,
and converts these to the same observables as the Q2B approach.
Table~\ref{tbl:AJBevan_rhopidata} summarises the experimental
constraints on $\B\to\rho\pi$ (DP analysis from \babar\ and Q2B
analysis from Belle).

\begin{table}[!h]
\begin{center}
\begin{tabular}{c|cccc}
mode & Expt & \S & \C & \acp \\ \hline

$B\to\pi^+\pi^-\pi^0$ & \babar & $-0.10\pm 0.14 \pm 0.04$ &
   $0.34\pm 0.11\pm 0.05$ & $-0.088\pm 0.049 \pm 0.013$ \\
$\Bztorhopi$ & Belle & $-0.28 \pm 0.23 ^{+0.10}_{-0.08}$ &
  $0.25 \pm 0.17^{+0.02}_{-0.06}$ & $-0.16 \pm 0.10 \pm 0.02$\\
\end{tabular}
\end{center}
\caption{Measurements of the $B \to \rho\rho$ decays.  BRs are in
units of $10^{-6}$.} \label{tbl:AJBevan_rhopidata}
\end{table}

\begin{figure}[!h]
\begin{center}
\resizebox{8cm}{!}{\includegraphics{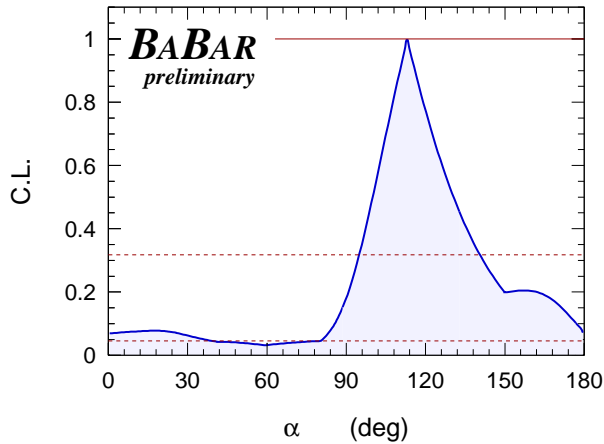}}
\end{center}
 \caption{A plot of the confidence level of $\alpha$ determined from \babar's $\rho\pi$ DP analysis.}
\label{fig:bevan_rhopi_isospin}
\end{figure}

Using SU(2) with \babar's DP result, one obtains the following
constraint; $\alpha=(113^{+27}_{-17} \pm 6)^\circ$ (Figure
\ref{fig:bevan_rhopi_isospin} shows the corresponding confidence
level plot for $\alpha$). This result is self consistent as the
strong phase differences and amplitudes are determined solely from
the structure of the DP. The unique aspect of this result is that
there is only a single solution between 0 and $180^\circ$, thus a
two fold ambiguity in the $\rho\eta$ plane. As a result this
measurement is an important constraint on alpha, ancillary to that
of the \babar\ $\Bztorhorho$ analysis.

It is possible to derive a model dependent constraint on alpha
from these data.  As shown by Gronau and Zupan~\cite{bevan:gronauzupan}.
One can make several assumptions, including (i) the relative
strong phase differences between tree level contributions for
$\rho^+\pi^-$ and $\rho^-\pi^+$ being less than $90^\circ$ (ii)
SU(3) being exact for penguin amplitudes (iii) neglecting
Electroweak penguins (iv) neglecting annihilation and exchange
diagrams to derive a constraint on $\alpha$. SU(3) breaking
effects are estimated in this model using CLEO, \babar\ and Belle
data. When doing this Belle obtain $\alpha=(102\pm 11_{expt} \pm
15_{model})^\circ$. Gronau and Zupan compute the corresponding
result for the \babar\ analysis as $(93\pm 4_{expt} \pm
15_{model})^\circ$.  There are other models proposed in the
literature, for example~\cite{bevan:gronaulunghi}.

\section{Comparison with the Standard Model}

The indirect measurements of $\alpha$ from the SM are $95\pm
9^\circ$~\cite{bevan:utfitter} and $98\pm 16^\circ$~\cite{bevan:ckmfitter}
from the UT Fit and CKM Fitter groups, respectively. Direct
measurements from the B-factories are in good agreement with these
predictions, where the average of $\alpha$ from $hh$ decays, using
isospin, is known to ${\cal O}(10^\circ)$. The precision of this
result is dominated by $B \to \rho\rho$, with an important
contribution from $B\to\rho\pi$, where the latter reduces this
constraint from four-fold to a two-fold ambiguity in the
$\rho\eta$ plane. In the next few years $\alpha$ should be known
to ${\cal O} (2-3^\circ)$ and the B-factories will be able to cleanly test
the closure of unitarity triangle with precision for the first
time.  A significant isospin analysis result from $B\to\pi\pi$ is
still some way off. Similarly, the model dependent calculations of
$\alpha$ are in agreement with the SM.

\section{Summary}

In the past year the B-factories have made significant progress
towards a precision measurement of the unitarity triangle angle
$\alpha$, only relying on SU(2) symmetry.  The dominant constraint
comes from $B\to \rho\rho$ decays~\cite{bevan:my_rhorhoprl}, which has a
four-fold ambiguity in the $\rho\eta$ plane. The \babar\ analysis
of $B \to \pi^+ \pi^- \pi^0$ decays represents the first
time-dependent \CP analysis. Inclusion of the interference regions
of the DP enables one to measure strong phase differences between
amplitudes contributing to the final state; In turn this enables
one to derive a self consistent constraint of $\alpha$ with  a two
fold ambiguity in the $\rho\eta$ plane.  One can also employ
models to further constrain $\alpha$ as has been discussed herein.

\subsection{Bibliography}

\section*{Acknowledgments}

The author wishes to thank M. Bona, A. H$\ddot{o}$cker, M. Perini
and J. Zupan for illuminating discussions in the preparation of
this work.  This work is supported in part by the U.S. Department of 
Energy under contract number DE-AC02-76SF00515 and by PPARC, UK.

%
\label{AJBevanEnd}

\end{document}